\documentclass[%
 aip,
 amsmath,amssymb,
 reprint,%
]{revtex4-1}

\usepackage{graphicx}
\usepackage{dcolumn}
\usepackage{bm}

\usepackage[utf8]{inputenc}
\usepackage[T1]{fontenc}
\usepackage{mathptmx}
\usepackage{etoolbox}
\usepackage{adjustbox}
\usepackage{hyperref}
\usepackage{cleveref}
\DeclareUnicodeCharacter{2212}{-}

\makeatletter
\def\@email#1#2{%
 \endgroup
 \patchcmd{\titleblock@produce}
  {\frontmatter@RRAPformat}
  {\frontmatter@RRAPformat{\produce@RRAP{*#1\href{mailto:#2}{#2}}}\frontmatter@RRAPformat}
  {}{}
}%
\makeatother
\begin{document}

\preprint{AIP/123-QED}

\title{Impact of recent updates to neutrino oscillation parameters on the effective Majorana neutrino mass in 0$\nu\beta\beta$ Decay}
\author{Dongming Mei}
 \email{dongming.mei@usd.edu}
\affiliation{ 
Physics Department, University of South Dakota, Vermillion, SD, 57069}
\author{Kunming Dong}
\affiliation{ 
Physics Department, University of South Dakota, Vermillion, SD, 57069}
\author{Austin Warren}
\affiliation{ 
Physics Department, University of South Dakota, Vermillion, SD, 57069}
\author{Sanjay Bhattarai}
\affiliation{ 
Physics Department, University of South Dakota, Vermillion, SD, 57069}
\date{\today}

\begin{abstract}
We investigate how recent updates to neutrino oscillation parameters and the sum of neutrino masses influence the sensitivity of neutrinoless double-beta (0$\nu\beta\beta$) decay experiments. Incorporating the latest cosmological constraints on the sum of neutrino masses and laboratory measurements on oscillations, we determine the sum of neutrino masses for both the normal hierarchy (NH) and the inverted hierarchy (IH). Our analysis reveals a narrow range for the sum of neutrino masses, approximately 0.06 eV/c$^2$ for NH and 0.102 eV/c$^2$ for IH. Utilizing these constraints, we calculate the effective Majorana masses for both NH and IH scenarios, establishing the corresponding allowed regions. Importantly, we find that the minimum neutrino mass is non-zero, as constrained by the current oscillation parameters. Additionally, we estimate the half-life of 0$\nu\beta\beta$ decay using these effective Majorana masses for both NH and IH. Our results suggest that upcoming ton-scale experiments will comprehensively explore the IH scenario, while 100-ton-scale experiments will effectively probe the parameter space for the NH scenario, provided the background index can achieve 1 event/kton-year in the region of interest.

\end{abstract}
\maketitle
\section{Introduction}
The revelation of neutrino oscillations in solar and atmospheric neutrino experiments \cite{SNO,Kamland,SuperK} has marked a significant leap in our comprehension of neutrinos, uncovering their mass and mixing within the lepton sector. This breakthrough holds profound implications for extending the standard model of particle physics and pursuing a grand unified theory of nature. Yet, despite this progress, numerous facets of neutrinos remain enigmatic, such as their absolute mass scale, mass hierarchy, and the determination of whether they exhibit a Dirac or Majorana nature~\cite{Maj}.

Various experiments, spanning accelerator-based and reactor-based neutrino oscillation studies, beta decay investigations, and neutrinoless double-beta ($0\nu\beta\beta$) decay research, contribute to our understanding of the absolute mass scale and mass hierarchy of neutrinos. However, it's noteworthy that $0\nu\beta\beta$ decay experiments uniquely offer a practical means to explore the Majorana nature of neutrinos. These experiments aim to address fundamental questions by probing the effective Majorana mass of electron neutrinos and determining the absolute neutrino mass scale, hierarchy, and character \cite{Ell02, Ell04, Avi04, Bar04}.

In a $0\nu\beta\beta$ decay experiment, the connection between the measured half-life (T$_{\beta\beta}^{0\nu}$) and the effective Majorana mass of the electron neutrino, $m_{\beta\beta}$, is approximated by the following expression~\cite{Avi08, Bar23}:
\begin{equation}
    \label{eq:decay1}
    (T_{\beta\beta}^{0\nu})^{-1} = G^{0\nu}(E_{0},Z)|(\frac{m_{\beta\beta}}{m_{e}})^2||M_{f}^{0\nu}-(\frac{g_{A}}{g_{V}})^2M_{GT}^{0\nu}|^2.
\end{equation}
In Equation \ref{eq:decay1}, $G^{0\nu}(E_{0}, Z)$ includes couplings and a phase space factor, where $g_{A}$ and $g_{V}$ represent the axial vector and vector coupling constants, and $M_{f}^{0\nu}$ and $M_{GT}^{0\nu}$ denote the Fermi and Gamow-Teller nuclear matrix elements, respectively. Consequently, the precision with which $m_{\beta\beta}$ can be determined from the measured half-life relies on the accuracy of the theoretical nuclear matrix elements.

If the $0\nu\beta\beta$ decay process is mediated by light Majorana neutrinos, the effective neutrino mass is determined by the following coherent sum:

\begin{equation}
\label{eq:mass1}
|m_{\beta\beta}| = \left| \sum_{i} m_{i}U_{\beta i}^{2} \right|,
\end{equation}

where $m_{i}$ represents the mass of the $i$th neutrino mass eigenstate, $U_{\beta i}$ are elements of the leptonic matrix $U$, and the sum extends over all light neutrino mass eigenstates~\cite{bpo}. In the standard three-flavor scheme, $U$ is a unitary matrix and can therefore be parameterized in terms of three flavor mixing angles and one CP-violating phase:

\begin{equation}
    \label{decay2}
   U = \left (
    \begin{array} {ccc}
     c_{12}c_{13} & s_{12}c_{13} & s_{13}e^{-i\delta} \\
     -s_{12}c_{23}-c_{12}s_{13}s_{23}e^{i\delta} & c_{12}c_{23}-s_{12}s_{13}s_{23}e^{i\delta}&c_{13}s_{23}\\
     s_{12}s_{23}-c_{12}s_{13}c_{23}e^{i\delta}&-c_{12}s_{23}-s_{12}s_{13}c_{23}e^{i\delta}&c_{13}c_{23}
    \end{array}
   \right )
\end{equation}
where $c_{ij} = cos\theta_{ij}$, $s_{ij} = sin\theta_{ij}$ (for ij = 12; 13; 23), $\delta$ represents the Dirac charge-parity (CP) phase. 

The effective Majorana mass is determined by the absolute value of the element $m_{\beta\beta}$ within the mass matrix in the charged lepton flavor basis. This mass matrix characterizes the charged leptons in their diagonal basis. Consequently, the Majorana neutrino mass matrix in the charged lepton flavor basis can be represented as:
\begin{equation}
    \label{eq:decay3}
    |m_{\beta\beta}| = |U \cdot D \cdot U^T|.
\end{equation}
In this context, the diagonal matrix can be denoted as $D = \text{diag}[m_1 e^{i\alpha}, m_2, m_3 e^{i\beta}]$, where both $\alpha$ and $\beta$ represent the Majorana CP-violating phases.

The results derived from Equation~\ref{eq:decay3} can be expressed as:
\begin{equation}
\label{eq:mass2}
|m_{\beta\beta}| = |m_{\beta\beta}^{(1)}e^{i\alpha} + m_{\beta\beta}^{(2)}+m_{\beta\beta}^{(3)}e^{i\beta}|,
\end{equation}
 and

\begin{equation}
\label{eq:mass3}
\begin{array}{ll}
|m_{\beta\beta}^{(1)}| = m_{1}c_{12}^{2}c_{13}^{2},\\
|m_{\beta\beta}^{(2)}| = m_{2}s_{12}^{2}c_{13}^{2},\\
|m_{\beta\beta}^{(3)}| = m_{3}s_{13}^{2}.
\end{array}
\end{equation}

After years of experimental exploration, significant progress has been made in determining the values of $\theta_{12}$, $\theta_{13}$, and $\theta_{23}$ with a high degree of accuracy from current neutrino oscillation data. However, the three phase parameters ($\delta$, $\alpha$, and $\beta$) remain elusive~\cite{pdg}. Additionally, while the value of $\Delta m_{21}^2 \equiv m_2^2 - m_1^2$ and the absolute value of $\Delta m_{31}^2 \equiv m_3^2 - m_1^2$ have been measured, the sign of $\Delta m_{31}^2$ and the absolute neutrino mass scale remain unknown.

Therefore, the magnitude of $|m_{\beta\beta}|$ is subject to three types of uncertainties, namely the unknown absolute neutrino mass scale ($m_1$ or $m_3$), the unknown neutrino mass hierarchy ($\Delta m_{31}^2 > 0$ or $\Delta m_{31}^2 < 0$), and the unknown Majorana phases $\alpha$ and $\beta$ appearing in $|m_{\beta\beta}|$, even in the absence of new physics contamination. Given that the magnitude of $|m_{\beta\beta}|$ is closely tied to the decay half-life of the $0\nu\beta\beta$ decay process, it is of significant interest to explore how these three uncertainties will influence its value. Evaluating this impact will provide valuable insights into the practical feasibility of conducting an experiment.

Until now, extensive phenomenological endeavors have been dedicated to exploring the parameter space of $|m_{\beta\beta}|$ and evaluating its sensitivity to potential new physics phenomena~\cite{ago}.
Current experiments have achieved remarkable sensitivities, with half-life measurements reaching levels of $\sim 10^{26}$ years for isotopes such as $^{136}$Xe and $^{76}$Ge, consequently setting upper limits on the effective Majorana neutrino mass within the range of 36-156 meV \cite{KamLandZen, Gerda}.

In this study, we aim to investigate the parameter space of $|m_{\beta\beta}|$ by incorporating the sum of neutrino masses constrained by recent cosmological data~\cite{cos} and utilizing the best-fitted oscillation parameters derived from global particle physics datasets~\cite{pdg}. Similar studies have been conducted by several authors~\cite{huang, raul, cao, bura}, including a recent one by Denton and Gehrlein~\cite{peter}, who focused on flavor models in the context of the funnel ($m_{\beta\beta} < 1$ meV) and explored related topics.

In our work, we will investigate several intriguing scenarios, including the allowed regions for the sum of neutrino masses and the minimum neutrino mass ($m_{1}$ or $m_{3}$). Additionally, we will analyze the behavior of $|m_{\beta\beta}|$ within these regions. Given an allowed sum of neutrino masses, we determine the value of the minimum neutrino mass. We point out that if the minimum neutrino mass falls outside the range of [2$\times10^{-3}$ eV/c$^2$, 7$\times10^{-3}$ eV/c$^2$], $|m_{\beta\beta}|$ will exceed 1 meV/c$^2$ in magnitude. This level of sensitivity can be achieved with a 100-ton experiment over 10 years. This observation elucidates the relationship between neutrino mass constraints and the effective Majorana mass. Utilizing the effective Majorana mass constrained by the minimum neutrino mass, we predict the sensitivity for a 100-ton experiment, highlighting that a background index of 1 event per kton per year is required to achieve an effective Majorana mass of 1 meV.

\section{the Sum of Neutrino Masses versus the minimum Neutrino Mass}

The most recent cosmological constraints on the sum of neutrino masses~\cite{cos}, coupled with the latest laboratory measurements on oscillations~\cite{pdg}, offer valuable insights into the constraints on the effective Majorana neutrino mass. 
In this work, we utilize the squared mass splitting constraints from a comprehensive global fit to neutrino oscillation observations, as presented in the first column of Table 14.7 from The Review of Particle Physics (2023) by the Particle Data Group~\cite{pdg}. Although there are three additional columns with slightly different fitted values in the same table, the differences are minimal and do not significantly impact the analysis. For the sake of simplicity, we employed only the parameters from the first column.

In the context of the normal neutrino mass hierarchy, often abbreviated as the normal hierarchy (NH), characterized by $m_1 < m_2 < m_3$, the parameters employed in this study are:
\begin{equation}
\label{eq:para1}
\left \{
 \begin{array} {c}
  \Delta m_{21}^2 = (7.41_{-0.20}^{+0.21}) \times 10^{-5} eV^2/c^4\\
  \\
  \Delta m_{32}^2 = (2.437_{-0.027}^{+0.028}) \times 10^{-3} eV^2/c^4 \\
  \\
  \theta_{13} = (8.54_{-0.12}^{+0.11})^o \\
  \\
   sin^2\theta_{13} = (2.203_{-0.059}^{+0.056}) \times 10^{-2} \\
   \\
   \theta_{23} = (49.1_{-1.3}^{+1.0})^o \\
   \\
   sin^2\theta_{23} = (5.71_{-0.23}^{+0.18}) \times 10^{-1} \\
   \\
   \theta_{12} = (33.41_{-0.72}^{+0.75})^o \\
   \\
   sin^2\theta_{12} = (3.03_{-0.11}^{+0.12}) \times 10^{-1} 
   \end{array}
   \right \}
   \end{equation}

  Similarly, in the case of the inverted neutrino mass hierarchy, commonly referred to as the inverted hierarchy (IH), characterized by $m_3 < m_1 < m_2$, the parameters utilized in this study are:

   \begin{equation}
\label{eq:para2}
\left \{
 \begin{array} {c}
  \Delta m_{21}^2 = (7.41_{-0.20}^{+0.21}) \times 10^{-5} eV^2/c^4\\
  \\
  \Delta m_{32}^2 = (-2.498_{-0.025}^{+0.032}) \times 10^{-3} eV^2/c^4 \\
  \\
  \theta_{13} = (8.57_{-0.11}^{+0.12})^o \\
  \\
   sin^2\theta_{13} = (2.219_{-0.057}^{+0.060}) \times 10^{-2} \\
   \\
   \theta_{23} = (49.5_{-1.2}^{+0.9})^o \\
   \\
   sin^2\theta_{23} = (5.78_{-0.21}^{+0.16}) \times 10^{-1} \\
   \\
   \theta_{12} = (33.41_{-0.72}^{+0.75})^o \\
   \\
   sin^2\theta_{12} = (3.03_{-0.11}^{+0.12}) \times 10^{-1} 
   \end{array}
   \right \}
   \end{equation}

Using the measured values of $\Delta m_{12}^2$ and $\Delta m_{32}^2$ in both the NH and IH cases, we can derive the sum $\Sigma = m_1 + m_2 + m_3$ as follows:

\begin{equation}
\label{eq:sum}
\left \{
\begin{array}{c}
   \Sigma = m_1 + \sqrt{(7.41_{-0.20}^{+0.21})\times10^{-5} + m_1^2} + \\
   \sqrt{(7.41_{-0.20}^{+0.21})\times10^{-5} + (2.437_{-0.027}^{+0.028})\times10^{-3} + m_1^2}, \\
   \end{array}
   \right \}
   \end{equation}
  and 
  \begin{equation}
\label{eq:sum1}
\left \{
\begin{array}{c}
   \Sigma = m_3 + \sqrt{(2.498_{-0.025}^{+0.032})\times10^{-3} + m_3^2} + \\\sqrt{(7.41_{-0.20}^{+0.021})\times10^{-5} + (2.498_{-0.025}^{+0.032})\times10^{-3} + m_3^2} \\
   \end{array}
   \right \}
   \end{equation}
The oscillation constraints outlined in Equations \ref{eq:sum} and \ref{eq:sum1} can be visualized in Figure 1, illustrating the permitted region in the $m_{L} - \Sigma$ plane. Here, $m_{L}$ denotes the minimum neutrino mass, representing $m_1$ for NH and $m_{3}$ for IH.
\begin{figure} [htbp]
  \centering
  \includegraphics[clip,width=0.9\linewidth]{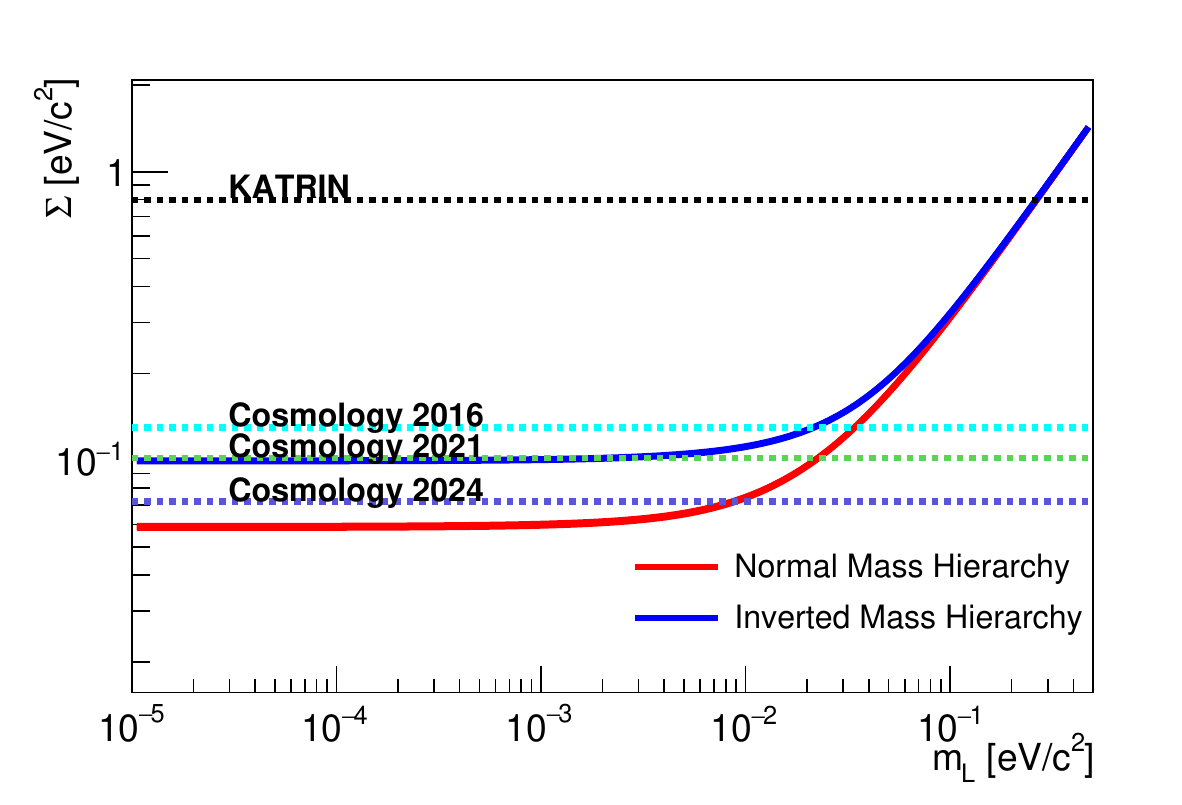}
  \caption{The sum of masses is represented by the symbol $\Sigma$, defined as $\Sigma = m_1 + m_2 + m_3$, and is depicted as a function of the minimum mass of neutrino mass eigenstates. The errors for each parameter stated in Equations~\ref{eq:para1} and \ref{eq:para2} are incorporated into the plot, but they are too small to have a significant impact. The estimated sensitivity of KATRIN is $\Sigma < 0.8$ eV (95\% C.L.) according to reference~\cite{kat}. Cosmology 2016 data is sourced from reference~\cite{ajc}, while Cosmology 2021 data is cited from reference~\cite{cos} and Cosmology 2024 is cited from reference~\cite{desi}.}
  \label{fig:f1}
\end{figure}

 As depicted in Figure~\ref{fig:f1}, the values of $\Sigma$ fall within two narrow bands: approximately $0.06$ eV/c$^2$ for NH and $\sim$0.102 eV/c$^2$ for IH. This suggests a minor influence of the value of the minimum neutrino mass when $m_{L} < 10^{-2} eV/c^2$. 
 For the case of IH, cosmological data from 2016 with a conference level of 95\% sets the upper limit of $m_L < 1.8 \times 10^{-2}$ eV/c$^2$. The 2021 data, also with a 95\% conference level, sets an upper limit of $8 \times 10^{-3}$ eV/c$^2$, and the 2024 data with a 95\% conference level has exceeded the IH scenario. In the case of NH, when combined with cosmological constraints, all with a 95\% conferences level, $m_L$ is less than $3.5 \times 10^{-2}$ eV/c$^2$ according to the 2016 data, less than $2 \times 10^{-2}$ eV/c$^2$ according to the 2021 data, and less than $9 \times 10^{-3}$ eV/c$^2$ according to the 2024 data. All of these results suggest that the quasidegenerate (QD) scenario is ruled out.
 It is noteworthy that the Particle Data Group~\cite{pdg} in 2023 provided three additional sets of best-fit parameters. Despite slight differences among these parameters, they do not alter the conclusion regarding the narrow allowed bands for $\Sigma$ in both NH and IH cases.

Recent searches in cosmology and accelerator experiments indicate a preference for the NH scenario for neutrino mass~\cite{raul, nova, t2k}. This implies that the dominant components of $|m_{\beta\beta}|$ are the lighter neutrino mass eigenstates, leading to longer decay half-lives and greater experimental challenges. This raises potential concerns for the planned ton-scale 0$\nu\beta\beta$ decay experiments, which aim to explore the IH scenario.  While the reliability of the cosmological claim has faced scrutiny~\cite{sch}, recent results from experiments such as NOvA~\cite{nova} and T2K~\cite{t2k} lack statistical significance. Interestingly, a combined NOvA and T2K analysis favors the inverted hierarchy (IH)~\cite{denton, kevin}. Additionally, by combining electron neutrino ($\nu_{e}$) disappearance data from reactor experiments and muon neutrino ($\nu_{\mu}$) disappearance data from accelerator-based experiments, the mass hierarchy can be determined~\cite{hiro, step}, providing further evidence supporting the NH. Considering all of the above, the situation remains quite complex at this stage. Therefore, more experiments are needed to conclusively determine the mass hierarchy. The planned ton-scale $0\nu\beta\beta$ experiments will definitively address the neutrino mass hierarchy and validate the cosmological models used to obtain constraints on the sum of neutrino masses.

As an example, to address the potential preference for the NH using $0\nu\beta\beta$ decay experiments, Agostini, Benato, and Detwiler \cite{ago} found that the likelihood of detecting $0\nu\beta\beta$ decay could exceed 50\% in the most promising experiments if there are no neutrino mass mechanisms that drive the mass of the lightest state or the effective Majorana mass to zero. Additionally, the possibility of decay facilitated by the exchange of heavy particles \cite{petr, shao}, which involves Majorana neutrinos according to the Black-Box Theorem \cite{jsch}, also violates lepton number conservation. However, determining the mass scale of the light neutrinos directly from the data in this scenario remains challenging.

While the implications of the NH based on limited experimental evidence are relatively weak, the potential for exotic heavy particle exchange physics could diminish their significance. Nevertheless, the primary motivation for investigating $0\nu\beta\beta$ decay—testing lepton number conservation and distinguishing between the Dirac or Majorana nature of neutrinos—remains compelling.

Given the narrow allowed regions of $\Sigma$ in both NH and IH cases, we aim to explore the parameter space of $|m_{\beta\beta}|$ versus the minimum neutrino mass $m_{L}$ for both scenarios, particularly when accounting for the Majorana CP-violating phases.

\section{The parameter Space of $|m_{\beta\beta}|$}
In the early 2000s, pioneers developed formulations to calculate 
$|m_{\beta\beta}|$ as a function of the minimum neutrino mass~\cite{fps}. Recently,
the coupling-rod diagram of $|m_{\beta\beta}|$, introduced by Zhi-Zhong Xing and Ye-Ling Zhou~\cite{xing}, illustrates the key characteristics of $|m_{\beta\beta}|$ in the complex plane. This diagram provides an intuitive understanding of how the neutrino mass ordering and CP-violating phases impact $|m_{\beta\beta}|$. Below, we present a summary of the maximum and minimum values of $|m_{\beta\beta}|$ using Equation~\ref{eq:mass2} for both NH and IH cases.

For the case of NH where $m_1 < m_2 < m_3$, the equations are: 
\begin{equation}
    \label{eq:NH}
    \left \{
    \begin{array}{c}
    |m_{\beta\beta}|_{max} = \sqrt{\Delta m_{21}^2 + m_1^2}sin^2\theta_{12}cos^2\theta_{13}[1+\\
    \sqrt{1-\frac{\Delta m_{21}^2}{\Delta m_{21}^2 + m_1^2}}cot^2\theta_{12} +\\
    \sqrt{1-\frac{\Delta m_{21}^2}{\Delta m_{21}^2 + m_1^2}+\frac{\Delta m_{21}^2+\Delta m_{32}^2}{\Delta m_{21}^2 + m_1^2}}\frac{tan^2\theta_{13}}{sin^2\theta_{12}}]. \\
    \\
    |m_{\beta\beta}|_{min}^{(1)} = \sqrt{\Delta m_{21}^2 + m_1^2}sin^2\theta_{12}cos^2\theta_{13}[1 - \\
    \sqrt{1-\frac{\Delta m_{21}^2}{\Delta m_{21}^2 + m_1^2}}cot^2\theta_{12} - \\
    \sqrt{1-\frac{\Delta m_{21}^2}{\Delta m_{21}^2 + m_1^2}+\frac{\Delta m_{21}^2 + \Delta m_{32}^2}{\Delta m_{21}^2 + m_1^2}}\frac{tan^2\theta_{13}}{sin^2\theta_{12}}]. \\
    \\
|m_{\beta\beta}|_{min}^{(2)} = \sqrt{\Delta m_{21}^2 + m_1^2}sin^2\theta_{12}cos^2\theta_{13}[\sqrt{1-\frac{\Delta m_{21}^2}{\Delta m_{21}^2 + m_1^2}}\times \\
cot^2\theta_{12} - 1-\\
    \sqrt{1-\frac{\Delta m_{21}^2}{\Delta m_{21}^2 + m_1^2}+\frac{\Delta m_{21}^2 + \Delta m_{32}^2}{\Delta m_{21}^2 + m_1^2}}\frac{tan^2\theta_{13}}{sin^2\theta_{12}}]. \\
    
    \end{array}
    \right \}
\end{equation}

For the IH case, where $m_3 < m_1 < m_2$, the equations are as follows:
\begin{equation}
    \label{eq:IH}
    \left \{
    \begin{array}{c}
    |m_{\beta\beta}|_{max} = \sqrt{m_3^2 - \Delta m_{32}^2}sin^2\theta_{12}cos^2\theta_{13}[1+\\
    \sqrt{1-\frac{\Delta m_{21}^2}{m_3^2 - \Delta m_{32}^2}}cot^2\theta_{12} +\\
    \sqrt{1-\frac{\Delta m_{21}^2}{m_3^2 -\Delta m_{32}^2}+\frac{\Delta m_{21}^2 - \Delta m_{32}^2}{m_3^2 - \Delta m_{32}^2 }}\frac{tan^2\theta_{13}}{sin^2\theta_{12}}]. \\
    \\
|m_{\beta\beta}|_{min} = \sqrt{m_3^2 - \Delta m_{21}^2}sin^2\theta_{12}cos^2\theta_{13}[
    \sqrt{1-\frac{\Delta m_{21}^2}{m_3^2 - \Delta m_{32}^2 }}\times \\cot^2\theta_{12} - 
    1- \\
    \sqrt{1-\frac{\Delta m_{21}^2}{m_3^2 - \Delta m_{32}^2 }+\frac{\Delta m_{21}^2 - \Delta m_{32}^2}{m_3^2 - \Delta m_{32}^2 }}\frac{tan^2\theta_{13}}{sin^2\theta_{12}}]. \\
    
    \end{array}
    \right \}
\end{equation}

\begin{figure} [htbp]
  \centering
  \includegraphics[clip,width=0.99\linewidth]{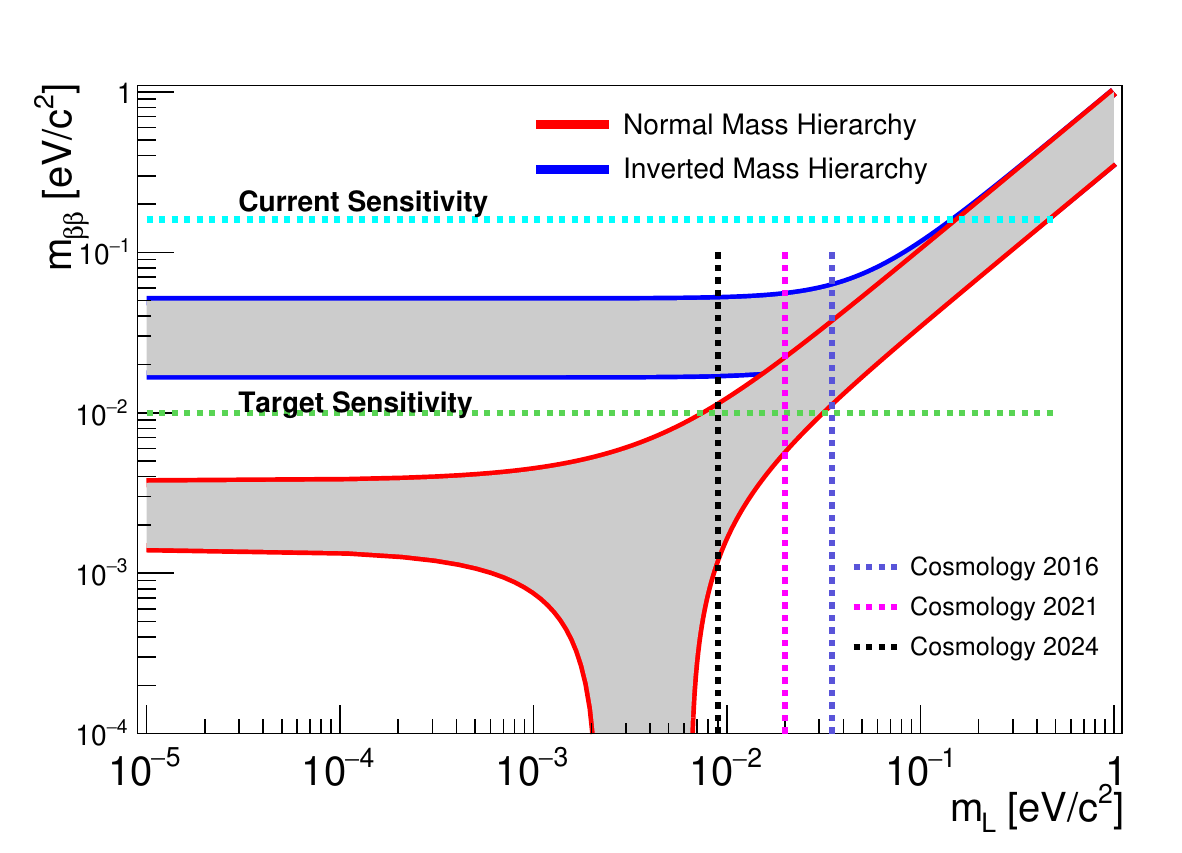}
  \caption{The variation of $|m_{\beta\beta}|$ with $m_{L}$ in the NH or IH of three neutrino masses is depicted. The input parameters from equations~\ref{eq:para1} and \ref{eq:para2} are utilized, with the relevant CP-violating phases allowed to range from 0 to $2\pi$. Note that the errors for each parameter have been evaluated. However, these errors are too small to produce a significant impact. Additionally, two horizontal lines represent the current sensitivity from KamLAND-Zen and GERDA, and the target sensitivity from planned ton-scale experiments such as LEGEND-1000 \cite{legend}, nEXO \cite{nexo}, and CUPID \cite{cupid}. The three vertical lines indicate the upper limits of $m_L$ constrained by cosmological data \cite{ajc, cos, desi}. }
  \label{fig:f2}
\end{figure}

We illustrate the relationship between $|m_{\beta\beta}|$ and the minimum neutrino mass $m_{L}$ in Figure~\ref{fig:f2}, utilizing the parameters specified in equations~\ref{eq:para1} and \ref{eq:para2}. We vary the relevant CP-violating phases within the range of 0 to $2\pi$.

In Figure~\ref{fig:f2}, the gray area represents the allowed regions for $|m_{\beta\beta}|$ concerning $m_{L}$. In the IH scenario, when $m_{L}$ is below $1 \times 10^{-2}$ eV/c$^2$, the allowed region for $|m_{\beta\beta}|$ appears nearly flat, ranging from 10 meV/c$^2$ to 50 meV/c$^2$. This region is targeted by upcoming ton-scale experiments like LEGEND-1000, nEXO, and CUPID. Conversely, in the NH scenario, the allowed region exhibits a pronounced dependence on $m_{L}$. For $m_{L}$ below $\sim$5$\times$10$^{-4}$ eV/c$^2$, the $|m_{\beta\beta}|$ region remains almost flat, surpassing 1 meV/c$^2$. However, within the range [$\sim2\times10^{-3}$, $\sim7\times10^{-3}$] eV/c$^2$, $|m_{\beta\beta}|$ tends towards zero. 

It's evident from Figure~\ref{fig:f2} that the planned ton-scale experiments will comprehensively explore the IH region. Therefore, our focus will be on accessing sensitivity for the NH case. We aim to investigate the feasibility of constructing a sensitive experiment capable of detecting the 0$\nu\beta\beta$ decay process if neutrino masses adhere to normal ordering.

\section{Discussion on $m_{L}$ and Prospective Experiments Sensitive to the NH Region}

To investigate the NH region for the $0\nu\beta\beta$ decay process, it is important to note that the effective Majorana mass can approach zero if $m_{L}$ falls within the range of [$\sim2\times10^{-3}$, $\sim7\times10^{-3}$] eV/c$^2$, as indicated in Figure~\ref{fig:f2}. Considering this, our first step is to examine the value of $m_{L} = m_1$ for a given allowed $\Sigma$. The solution must satisfy the following condition:
\begin{equation}
    \label{eq:e1}
    m_L + \sqrt{\Delta m_{21}^2 + m_{L}^2} + \sqrt{\Delta m_{32}^2 + \Delta m_{21}^2 + m_{L}^2} - \Sigma = 0.
\end{equation}

\begin{figure} [htbp]
  \centering
  \includegraphics[clip,width=0.99\linewidth]{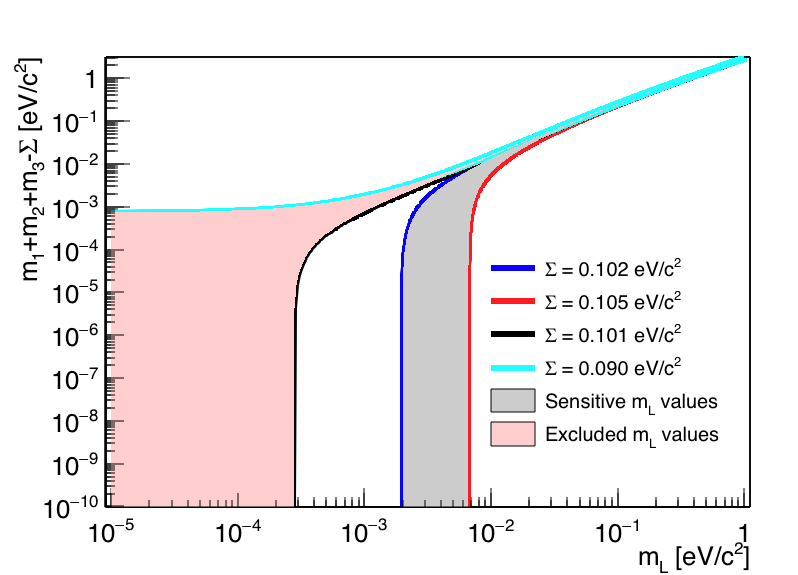}
  \caption{The plot illustrates the solution for $m_1+m_2+m_3$-$\Sigma$ = 0 when a $\Sigma$ value is specified. The gray area represents where the value of $|m_{\beta\beta}|$ approaches zero, while the light-red region indicates excluded values of $m_{L}$ due to the absence of a solution. Note that the errors for each parameter shown in Equations~\ref{eq:para1}
and \ref{eq:para2} are too small to have a significant impact on the outcomes.}
  \label{fig:f3}
\end{figure}

Equation~\ref{eq:e1} must be solved numerically for a given value of $\Sigma$. Figure~\ref{fig:f3} illustrates the relationship between $m_{L}$ and a given $\Sigma$.

Numerical solutions indicate that when $\Sigma$ ranges between 0.059 and 0.061 eV/c$^2$, Equation~\ref{eq:e1} is satisfied. This corresponds to $m_{L}$ falling within the region [3$\times10^{-4}$, 2$\times10^{-3}$] eV/c$^2$. However, when $\Sigma$ = 0.058 eV/c$^2$, Equation~\ref{eq:e1} is violated, indicating no suitable values of $m_{L}$ exist and thus excluding that range. Furthermore, for $\Sigma$ in the range [0.061, 0.068] eV/c$^2$, $m_{L}$ drives $|m_{\beta\beta}|$ to zero, as depicted in Figure~\ref{fig:f2}.

\begin{figure} [htbp]
  \centering
  \includegraphics[clip,width=0.99\linewidth]{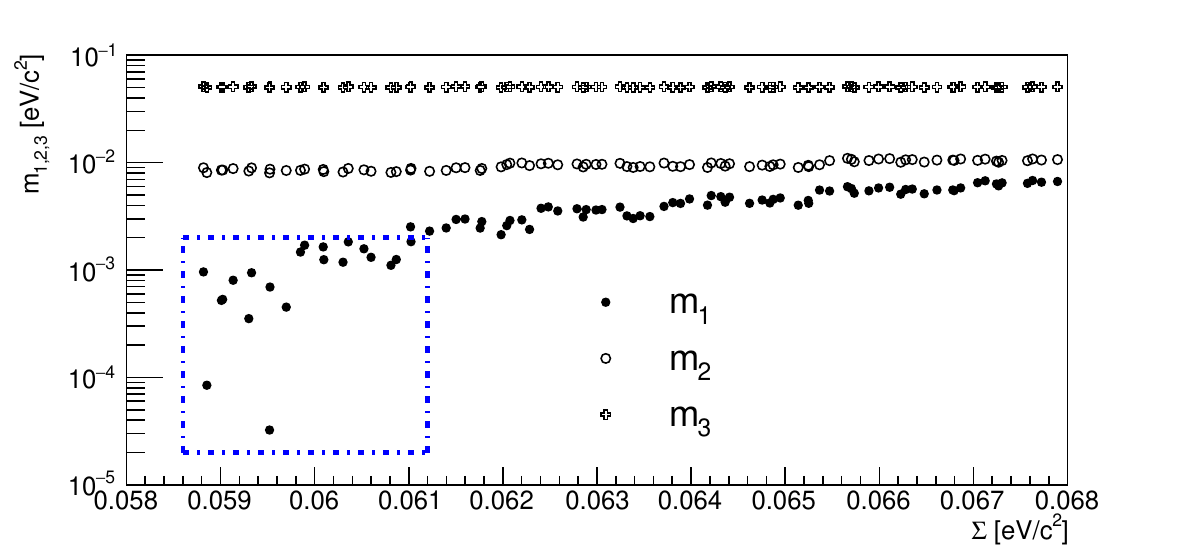}
  \caption{Displayed is the solution for the neutrino masses $m_1, m_2$, and $m_3$ for a given allowed $\Sigma$. The box outlined by dashed blue lines indicates the potential non-zero values of $m_1$. Note that the errors for each parameter shown in Equations~\ref{eq:para1}
and \ref{eq:para2} are incorporated into the calculation. However, they are too small to have a significant impact on the outcomes. }
  \label{fig:f4}
\end{figure}

It's evident that only a narrow range of $m_{L}$ is viable for determining $|m_{\beta\beta}|$ in future experiments beyond the ton-scale. This viable region spans $m_{L}$ = [3$\times10^{-4}$, 2$\times10^{-3}$] eV/c$^2$. Additionally, the corresponding of $m_1$, $m_2$, and $m_3$ for a given $\Sigma$ is visualized in Figure~\ref{fig:f4}. 

To be clear, in Figure~\ref{fig:f4}, for each $\Sigma$, the corresponding values of $m_1$, $m_2$, and $m_3$ are plotted. This indicates a unique solution for $m_1$, $m_2$, and $m_3$ corresponding to each given $\Sigma$. In the NH case, the values of $m_1$ within the box highlight the solution for each given $\Sigma$ in the allowed region shown in Figure~\ref{fig:f3}.

In the NH scenario, it's noteworthy that the sum of masses falls within the range of 0.059 to 0.061 eV/c$^2$, consequently leading to a non-zero $m_1$, as depicted in Figure~\ref{fig:f4}. This allowed region is consistent with the cosmological constraints from the 2016, 2021, and 2024 data analyses.

All the results depicted in Figures~\ref{fig:f3} and \ref{fig:f4} indicate that the minimum neutrino mass is non-zero. This suggests that only the Majorana CP-violating phases can drive $|m_{\beta\beta}|$ to zero when $m_{L}$ falls within the range of $[2 \times 10^{-3}, 7 \times 10^{-3}]$ eV/c$^2$. There exists a parameter space for $|m_{\beta\beta}|$ where its value is in the region of $\sim$ 1 meV/c$^2$, corresponding to the sensitivity achievable with a 100-ton-scale experiment. 

Using $^{76}$Ge as an example, the sensitivity to the effective neutrino mass is linked to the upper limit of the half-life, determined as:
\begin{equation}
\label{eq:e2}
T_{\frac{1}{2}} =  4.17 \times 10^{26}  (\frac{\varepsilon \cdot a}{W}) \frac{M \cdot t}{\mu},
\end{equation}
where $T_{\frac{1}{2}}$ is the decay half-life in years, $\varepsilon$ is the detection efficiency, $a$ is the isotopic abundance (92\% for $^{76}Ge$), $W$ is the molar mass of the source material, $M$ is the total mass of the source in kg, $t$ is the run time in years, and $\mu$ is the Poisson signal mean constrained by the number of background events ($B$) in the region of interest (ROI). We express $B$ as:

\begin{equation}
\label{eq:back}
B = M \cdot t \cdot b \cdot \Delta E,
\end{equation}
where $b$ is the number of background events per kg per keV per year, and $\Delta E = 3.0$ keV is the width of the ROI.

A 90\% confidence level interval for the Poisson signal mean ($\mu$) is provided in TABLE XII of Ref.~\cite{Gary} when the total number of events observed ($n_{0}$) and the mean background ($b$) are known. 

\begin{figure} [htbp]
  \centering
  \includegraphics[clip,width=0.99\linewidth]{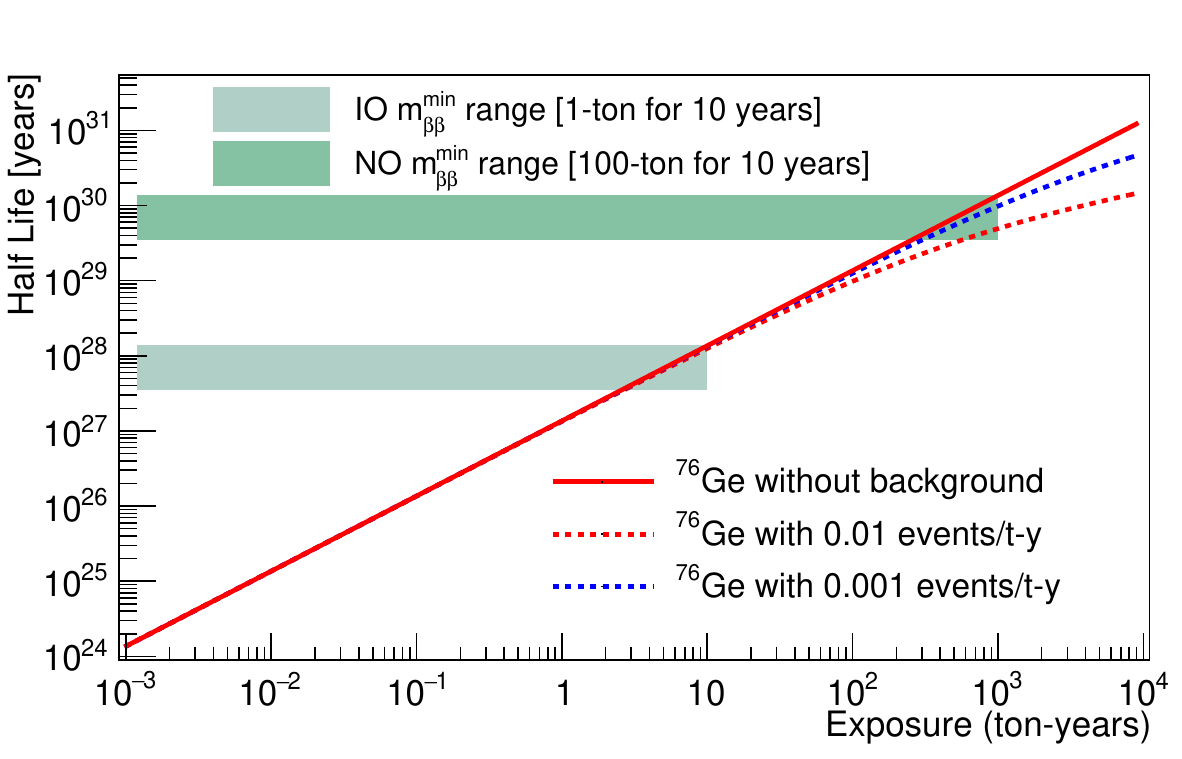}
  \caption{ The sensitivity for measuring the half-life of $^{76}$Ge decay is illustrated for a future 100-ton scale experiment utilizing similar technology to LEGEND-1000. The length and width of the shaded regions represent the assumed exposure and the allowed $t_{1/2}$ space, under the assumption that the decay is mediated by the light Majorana neutrino.}
  \label{fig:f5}
\end{figure}

Using equation~\ref{eq:e2} and parameters similar to those of LEGEND-1000~\cite{legend}, we demonstrate the sensitivity of such an experiment in Figure~\ref{fig:f5}.  Achieving a half-life of $10^{30}$ years with a 10-year exposure requires a background index of 0.001 events/t-y, presenting a considerable challenge. This underscores the necessity for significant research and development (R\&D) efforts to be initiated as early as possible. The development of $0\nu\beta\beta$ decay experiments has taken decades for the community to reach the current stage for the planned ton-scale experiments that will address the IH scenario. Historically, there was a claim for evidence of $0\nu\beta\beta$ decay from a subgroup of the Heidelberg-Moscow experiment~\cite{kkdc} using the enriched $^{76}$Ge. Despite this claim being refuted by GERDA~\cite{Gerda} and Majorana Demonstrator (MJD)~\cite{MJD} using the same isotope, the background index was 0.113 events/(kg$\cdot$keV$\cdot$year) in 2001. GERDA and MJD improved the background index by nearly a factor of 100 over a 17-year period from 2001 to $\sim$2017. These improvements included the use of better radio-pure materials for constructing experiments, better detectors with enhanced energy resolution and pulse shape analysis, active veto systems for GERDA, and electroformed copper for the inner shield in MJD. The planned LEGEND-1000 has shown the capacity to improve the background index by another factor of 100 with a ton-scale detector before 2030, incorporating active veto systems with underground argon. From 2017 to $\sim$2030 is another decade-long effort. To continue improving the background index by a factor of 30 for a 100-ton experiment will require significant effort, time, and innovative ideas, such as further enhancements in energy resolution and signal identification with machine learning and the use of large-size detectors to further reduce backgrounds, complexity, and cost.

\begin{figure} [htbp]
  \centering
  \includegraphics[clip,width=0.99\linewidth]{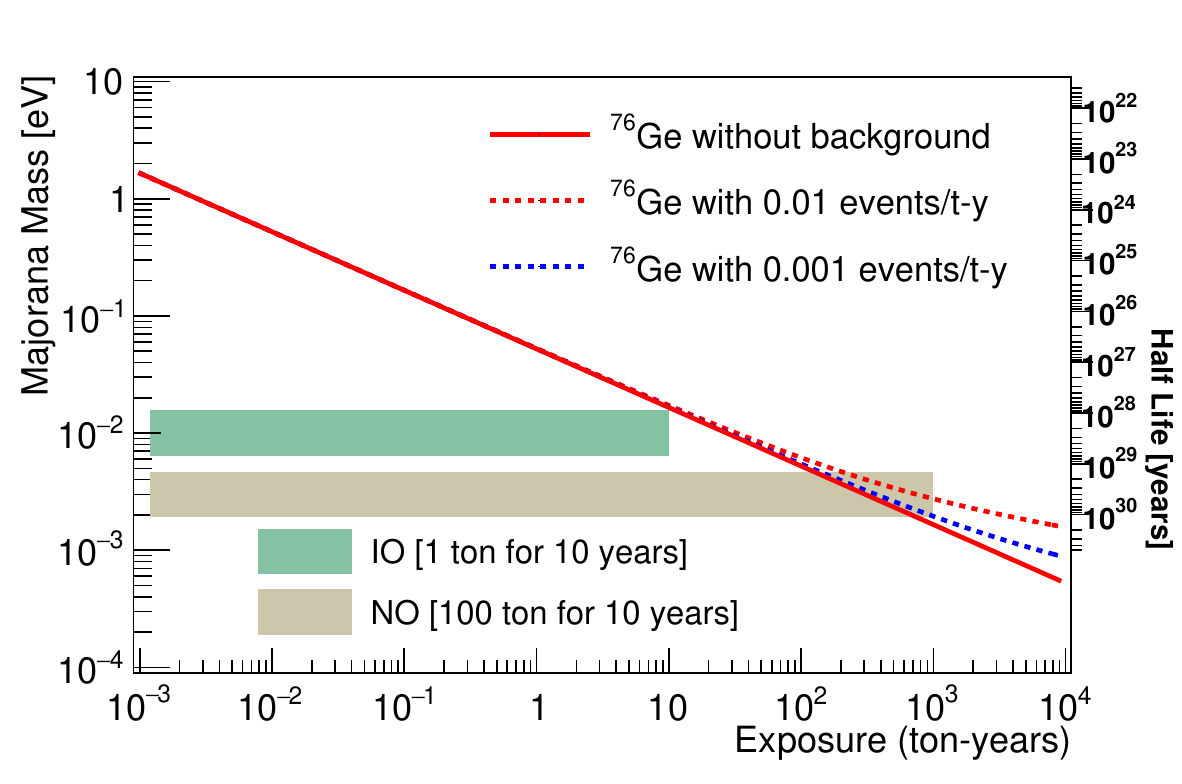}
  \caption{ Displayed is the sensitivity for a prospective 100-ton scale $^{76}$Ge experiment, leveraging technology akin to that of LEGEND-1000. The length and width of the shaded regions represent the assumed exposure and the allowed $m_{L}$ space, under the assumption that the decay is mediated by the light Majorana neutrino. }
  \label{fig:f6}
\end{figure}

Using equation~\ref{eq:decay1} and defining $F_{N} = G^{0\nu}(E_{0},Z)|M_{f}^{0\nu}-(\frac{g_{A}}{g_{V}})^2M_{GT}^{0\nu}|^2$, if a nuclear structure parameter $F_N = 7.01 \times 10^{14}/y$ is employed~\cite{34}, the effective neutrino mass $|m_{\beta\beta}|$ can be expressed in terms of the calculated $F_N$ and the measured half-life as:

\begin{equation}
    \label{eq:e3}
    |m_{\beta\beta}| = \frac{m_{e}}{\sqrt{F_{N}T_{1/2}^{0\nu}}},
\end{equation}
where $m_{e}$ is the electron mass. The measurable half-life for a 100-ton scale $^{76}$Ge-based experiment is depicted in Figure~\ref{fig:f6}.

As depicted in Figures~\ref{fig:f5} and \ref{fig:f6}, constructing a 100-ton scale experiment allows for comprehensive exploration of the $\sim$1 meV/c$^2$ region in the NH scenario, with the corresponding half-life approaching approximately $10^{30}$ years. This endeavor entails the preparation of enriched materials over a few years, alongside ongoing R\&D efforts aimed at continuously reducing backgrounds. Simultaneously, the scientific community is advancing towards ton-scale experiments to fully investigate the 10 meV/c$^2$ region in the IH scenario. Notably, the investigators are exploring the feasibility of dissolving $^{130}$Te or $^{136}$Xe isotopes into liquid scintillator to create a 100-ton scale experiment with JUNO aimed at exploring the $|m_{\beta\beta}| \approx 1$ meV/c$^2$ region~\cite{cao}. Additionally, they are contemplating the use of a Xe-doped liquid argon TPC as a platform for $0\nu\beta\beta$ decay with the DUNE far detectors, targeting a sensitivity of $|m_{\beta\beta}| \approx 2$ meV~\cite{dune}. Achieving sensitivity to $\sim 1$ meV in NH with a 100-ton scale experiment requires reducing the background to 1 event per kiloton per year in the ROI, as illustrated in Figures~\ref{fig:f5} and \ref{fig:f6}. Furthermore, exceptional energy resolution is essential for distinguishing events from 2$\nu\beta\beta$ decays. Both concepts proposed for JUNO and DUNE are intriguing but will face considerable challenges in meeting these critical requirements. For instance, both concepts will need to use extremely radio-pure materials for constructing the detectors, which is not a primary requirement for the main physics goals of JUNO and DUNE. Additionally, both experiments will need to significantly increase photon detection efficiency to improve their energy resolution.

\section{Conclusion}
Utilizing constraints from cosmology and neutrino oscillation experiments, we discuss the permissible parameter space of $|m_{\beta\beta}|$ for both the inverted hierarchy (IH) and normal hierarchy (NH) scenarios. Given the current planned ton-scale experiments aimed at addressing the sensitivity across the entire region of IH, we focus our discussion on the allowed regions of  $|m_{\beta\beta}|$ for NH. It becomes evident that the minimum neutrino mass ($m_{L}$) is non-zero in the NH case. To explore a non-zero $|m_{\beta\beta}|$, it is important to note that there is only a narrow region where the minimum neutrino mass lies between 3$\times10^{-4}$ to 2$\times10^{-3}$ eV/c$^2$ for NH. An achievable value of $|m_{\beta\beta}|$ is approximately 1 meV/c$^2$.
We elucidate the importance for future experiments to attain this sensitivity. With such heightened sensitivity, attainable through a high-capacity experiment (100-ton scale), the likelihood of observing signal events of 0$\nu\beta\beta$ decays substantially increases.

 \section{Acknowledgement} 
  This work was supported in part by NSF OISE 1743790, NSF PHYS 2310027, DOE DE-SC0024519, DE-SC0004768,  and a research center supported by the State of South Dakota.

\end{document}